\documentclass[12pt]{article}
\usepackage[cp1251]{inputenc}
\usepackage[T2A]{fontenc}
\usepackage[T1]{fontenc}
\usepackage{ae,aecompl}

\usepackage{latexsym,amsmath,amscd,amssymb,graphics}

\usepackage{mathrsfs} 
\DeclareMathAlphabet{\mathscrbf}{OMS}{mdugm}{b}{n} 
\usepackage{enumerate,framed}
\usepackage{graphicx}
\usepackage[colorlinks]{hyperref}

\evensidemargin\oddsidemargin \topmargin2cm

\usepackage{amsfonts,amscd,amsmath}

\newtheorem{theorem}{Theorem}

\newtheorem{definition}[theorem]{Definition}

\newtheorem{remark}[theorem]{Remark}

\begin{document}

\title{Randomized Quantum Hamiltonian Systems}
\date{}
\maketitle

\begin{center}
John E. Gough \footnote{Aberystwyth University, Wales},
Yurii N. Orlov\footnote{Keldysh Institute of Applied Mathematics RAS}, Vsevolod Zh. Sakbaev \footnote{Moscow Institute of Physics and Technology}, Oleg G. Smolyanov\footnote{Lomonosov Moscow State University, Moscow Institute of Physics and Technology}

\end{center}

\begin{abstract}
We present a procedure for averaging random unitary groups and random self-adjoint groups.
Central to this is a generalization of the notion of weak convergence of a sequence of measures and the corresponding
generalization of the concept of convergence in distribution. The convergence is established in
determination of the sequence of compositions of independent random transformations. When sequences of compositions of independent random transformations of the shift by the Euclidean vector in space, the results obtained coincide with the central limit theorem for the sums
independent random vectors. The results are applied to the dynamics of quantum systems arising
random quantization of the classical Hamiltonian system.
\end{abstract}

\section{Introduction}
Hamiltonian ensembles arise from defining a random variable $:\omega \mapsto H(\omega )$ on a probability space taking values in the self-adjoint operators on a fixed Hilbert space. It is of interest to then study the properties of the associated random unitary evolution generated by the random Hamiltonian $H$. Linear quantizations of Hamiltonian systems were introduced in \cite{Berezin} and developed in \cite{Orlov, Orlov20}. Here a randomization of a quantization of a Hamiltonian system may be seen as a map of the linear space of Hamiltonian functions defined on a symplectic phase space, which assigns to each Hamiltonian function a random variable whose values are self-adjoint operators acting in the Hilbert space of a quantum system. Our proposed scheme covers all known methods of quantization, since the value of a random variable at each element of the probability space describes some quantization. Moreover, the random semigroup generated by the random Hamiltonian is a random process with values in the group of unitary operators.

 Interest in the study of random variables with values in a set of unbounded operators in Banach spaces and random processes with values in a Banach space of bounded linear operators arises in a number of problems in mathematical physics \cite{Orlov20, VS14}. So, if in functional mechanics the Hamilton function can be a random variable \cite{Vo, VS14}, then the quantum analog of functional mechanics is a quantum system, the Hamiltonian of which is a random variable with values in the set of self-adjoint operators \cite{VS14, OSS-2016}.

The randomness of the Hamiltonian of a quantum system is a consequence of the ambiguity of the quantization procedure for classical Hamiltonian systems \cite{Berezin, OSS-2016}. As such, there are two sources of randomness in the study of the dynamics of quantum systems:

\textit{
1. The Hamiltonian of a classical system is a random variable with values in the space of continuously differentiable functions on the phase space;}

\textit{
2. The quantization of the classical Hamiltonian system is a random variable with values in the space of linear maps of the linear manifold of the space of smooth functions on the phase space into a linear manifold in the space of self-adjoint operators acting in the Hilbert space of a quantum system.}

It is of interest to study the influence of both each of the two sources of randomness and their joint influence on such characteristics of the quantum dynamics of a system as the average value of a random dynamic semigroup, its covariance characteristics, and probability distributions over the set of admissible quantum states.

The paper gives a definition of the weak convergence of a sequence of measures in terms of the convergence of a sequence of linear operators associated with these measures acting in a suitable topological vector space of test functions. It is shown that this definition includes the definition of weak convergence of measures if we choose the space of bounded continuous functions endowed with the topology of pointwise convergence as the space of test functions of functions. With the help of the introduced generalization of the weak convergence of measures, statements about the convergence in distribution of vector-valued random processes are obtained.

It is shown that Chernoff's theorem on the approximation of operator semigroups is a limit theorem describing the limit distributions of vector-valued random processes, and includes the central limit theorem as one of the realizations of the statement of Chernoff's theorem.

\subsection{Random operators and semigroups}
To study random operator-valued functions, we introduce the following terminology. We call a random variable, $\xi$, an $\mathscr{A}$-measurable function on a probability space $(\Omega ,\mathscr{A},\mathbb{P} )$ with values in some measurable space (specifically a topological space equipped with the Borel $\sigma$-algebra of subsets).

For example, if such a random variable takes values in the set of self-adjoint operators, then it is called a random Hamiltonian; the concept of a random semigroup is defined similarly (Definitions \ref{def:random_semigroup} and \ref{def:random_Hamiltonian}).

We fix a Banach space $X$ and consider the Banach space $B(X)$ of bounded linear transformations of $X$.
Let $Y_s(X)$ be the topological vector space of strongly continuous one-parameter mappings $F$ from the half-line $\mathbb{R}_+=[0,+\infty )$ to $B(X)$ with topology $\tau _s$ on defined by the family of functionals $\rho _{T,u},\, T\geq 0,\, u\in X$, given by $\rho _{T,u}(F)=\sup\limits_{t\in [0,T]}\|F(t)u\|_X$. In the following, we denote the pre-dual of $X$ by $X_*$, so that $(X_*)^*=X$. For $f\in X $ and $g \in X_\ast$, we denote the natural duality by $\langle g,f \rangle$.

\begin{definition}
\label{def:random_semigroup}
A random operator valued function is a random variable with values in a topological vector space $Y_s (X)$. A random semigroup $G$ is defined to be a random variable taking values in the set $S(X)$ of strongly continuous one-parameter semigroups of operators acting in the space $X$.
\end{definition}

The expectation of a random semigroup $G$ will be understood as the Pettis integral
$ \mathbb{E}[G] =\int\limits_{\Omega }G _{\omega } \, \mathbb{P} (d\omega  )$. That is,  $\mathbb{E}[G] $ is the element of $Y_s(X)$ such that
\begin{eqnarray}
\langle g,  \mathbb{E}[G] (t)f \rangle =\int\limits_{\Omega } \langle g, G _{\omega } (t )f  \rangle \mathbb{P} (d \omega),
\end{eqnarray}
for each $t\in \mathbb{R}_+,f\in X,g\in X_*$.

\begin{theorem}[\cite{OSS-2016}]
Let $(\Omega , \mathscr{A},\mathbb{P})$ be a probability space and consider a random variable $\xi$ with values in the space $Y_s(X)$ is uniformly bounded and densely strongly equicontinuous, i.e. there is a dense linear subspace $D\subset X$ such that for any $f\in D$ and any $\epsilon >0$ there is a number $\delta >0$ 
such that  $\sup\limits_{t\ge 0 ,\omega \in \Omega} \| \xi _{\omega }(t^\prime )f-\xi _{\omega }(t)f \|_X \leq \epsilon$ whenever $t^\prime\in \mathbb{R}_+$ with $|t^\prime - t | \leq \delta$.
Then $\mathbb{E}[ \xi ] \in Y_s(X)$.
\end{theorem}

\begin{theorem}[P. Chernoff, 1968,  \cite{Chernoff,Sinha_Srivastava}]
Let $X$ be a Banach space and take $F:\
[0,+\infty )\, \to \, B(X)$ to be continuous is the strong operator topology
with $F(0)=I$ and satisfying the inequality $ \|F(t)
 \Vert_{B(X)}\leq e^{\alpha t}, t\geq 0,$ for some $\alpha\geq 0$. If $F'(0)$ is closable and its closure is the generator of a strongly continuous semigroup of operators $U(t), t\geq 0$, the for all $u\in X$ and $T >0$ we have
\begin{eqnarray}
\lim\limits_{n\to \infty }\sup\limits_{t\in [0,T]}\|U(t)u-\left(F(\frac{t}{n})\right)^nu\|_X=0.
\end{eqnarray}
\end{theorem}

Let $\Pi (X)$ denote the set of strongly continuous mappings $F$ from $\mathbb{R}_+$ to $B(X)$ satisfying $F(0)=I$. 

\begin{definition}[\cite{OSS-2016}]
We say that operator functions $F,\,G\in \Pi (X)$ are Chernoff equivalent if, for each $T>0$ and any $u\in X$, the following condition is satisfied
\begin{eqnarray}
\lim\limits_{n\to \infty }\sup\limits_{t\in [0,T]}\|\left(G(\frac{t}{n})\right)^n u-\left(F(\frac{t}{n})\right)^nu\|=0
\end{eqnarray}
\end{definition}

\begin{definition}[\cite{OSS-2016}]
The generalized expectation of a random semigroup of operators $\xi$ is the one-parameter $U$, which is Chernoff equivalent to the expectation $\mathbb{E}[\xi ]$.
\end{definition}

\begin{theorem}[\cite{OSS-2017}]
Let  $H $ be a random variable taking values in the self-adjoint operators on a Hilbert space $\mathfrak{h}$ and let $U(\omega ,t)=\exp(i H (\omega )t),\, t\geq 0,\, \omega \in \Omega ,$ be the corresponding semigroup.
Let $\mathcal D$ be a dense subspace of $\mathfrak{h}$ such that $\int_{\Omega }\|H (\omega )u\|_\mathfrak{h} \, \mathbb{P} (d\omega )<\infty $ for each $u\in \mathcal D$.
Then, if the operator on $\mathcal D $ defined by $\overline{H} u=\int\limits_{\Omega } H(\omega )u \, \mathbb{P} (d\omega)$ is essentially self-adjoint, then the average $\overline{ F}(t)=\int\limits_{\Omega }U(\omega ,t) \mathbb{P} (d\omega ),\, t\geq 0$ is Chernoff equivalent to the semigroup $\overline{U}=e^{it\overline{ H}},\, t\geq 0$.
\end{theorem}

If a random semigroup $G$ is defined on a probability space $(\Omega,\mathscr{A},\mathbb{P} )$, then its generator $H_G$  is a random variable on the same probability space, determined by the condition: for each $\omega \in \Omega$, the random variable $H_G (\omega )$ is a generator of the semigroup $G( \omega )$. The random variable $H_G$ takes on a value in the set of generators $G(X)$ of strongly continuous one-parameter semigroups acting in the space $X$. The topology $\tau_G$ on the set $G(X)$ is determined by the condition that the bijection $J$ between $S(X)$ and $G(X)$, for which each semigroup corresponds to its generator, is a homeomorphism of topological spaces and $(S(X),\tau _{S})$ and $({\cal G}(X),\tau _{\cal G})$. In this way,  $H_G ={\cal J}\circ G $.

\begin{definition}
\label{def:random_Hamiltonian}
A random Hamiltonian is a measurable function on a probability space $(\Omega,\mathscr{A},\mathbb{P} )$ that takes values in the topological space of generators $({\cal G}(X),\, \tau _{\cal G})$. If the expectation of a random semigroup $G$ is an operator-valued function $F_G$ that is Chernoff equivalent to some semigroup $U_G$, then the generator $H_G$ of the semigroup is called the Feynman-Chernoff average of the random generator of the random semigroup $G$.
\end{definition}
  
\subsection{Linear Quantization Method}
As a motivation for the development of the analysis of random operators, we consider the quantization procedure as a random variable with values in the set of self-adjoint operators. There are various definitions of quantizing such a Hamiltonian system. We consider symmetric linear quantizations $Q_{\omega},\, \omega \in \Omega$, (see \cite{Berezin, Orlov}), each of which associates the Hamilton function $H_{\omega}=Q_{\omega }(h)$ with a self-adjoint operator in a Hilbert space $E$.

\begin{definition}
Specifying the structure of a probability space $(\Omega ,\mathscr{A},P)$ on a set of linear quantizations is called random linear quantization.
\end{definition}

Random linear quantization allows us to consider the quantum Hamiltonian of the system as a random self-adjoint operator, and the family of unitary groups, $\exp (-itH_{\omega} ) ,\, t\in \mathbb{R} ,\ \omega \in \Omega $, as a random group, provided that for all $t \in \mathbb{R}$ the mapping is a measurable mapping $\omega \to \exp (-itH_{\omega })$ with respect to the weak operator topology $\Omega \to B(\mathfrak{h})$. The last condition is satisfied if the space $\Omega$ is discrete.

Consider an example when a random generator can take only two values $ H_1$ and $ H_2$, both Hermitean operators (Hamiltonians), with probabilities $p_1$ and $p_2$. Here the Hamiltonians) generate semigroups $ U_1(t)=e^{t H_1},\, t\geq 0$ and $ U_2(t)=e^{t H_2},\, t\geq 0$, for $t \ge 0 $.
Consider the operator $ \overline{H}=p_1 H_1+p_2 H_2,\ p_1\geq 0,\, p_2\geq 0,\, p_1+p_2=1$.
This situation arises, for example, when the form of the quantum operator depends on the quantization rule. This is, in particular, a typical situation for linear quantization (see \cite{Vo}), when the Hamiltonian is a linear combination of Hermitian operators obtained according to different quantization rules, for example, according to Jordan and Weyl. Let the operators $ H_1$ and $ H_2$ be such that the conditions of Theorem 1 are satisfied. Then, in accordance with formula (2), the semigroup, $ \overline{U } (t) = e^{t \overline{H} }$, is Chernoff equivalent to a linear combination $p_1 U_1+p_2 U_2$ of the indicated semigroups.

\subsection{Limit Theorems}
Let $\xi _j,\, j\in \mathbb N$, be a sequence of independent identically distributed random variables with values in the Banach space $B(\mathfrak{h})$ of bounded linear operators in the Hilbert space $\mathfrak{h}$. The law of large numbers characterizes the limiting distribution of the averaged random variable, $ \overline \xi_n=\frac{1}{ n}\xi _n+...+ \frac{1}{ n}\xi _1$, and the central limit theorem characterizes the limiting distribution of the random variable $\hat \xi
_n=\frac{1}{ {\sqrt n}}\xi _n+...+ \frac{1}{ {\sqrt n}}\xi _1$.  Let us investigate the asymptotics for probability distributions of a sequence of random variables representing some averaged compositions of operator-valued random variables. For example, in the study of the law of large numbers for the composition of independent identically distributed random semigroups, the asymptotics of the distributions of the averaged random variable $\overline \xi _n=(\xi _n)^\frac{1}{ n}\circ...\circ (\xi _1)^\frac{1}{ n}$ was investigated (see \cite{OSS19}).

\begin{definition}
Let $\big( U_n(t),\, t\geq 0\big)$ be a sequence of independent random semigroups. Then the sequence satisfies the law of large numbers in the strong operator topology if
\begin{eqnarray}
\lim\limits_{n\to \infty}  \mathbb{P} \bigg(  
\| U_n(\frac{t}{n})\circ ... \circ U_1(\frac{t}{n}) \, x -
\overline{U_n(\frac{t}{n})\circ ... \circ U_1(\frac{t}{n})}
 \, x\|_\mathfrak{h}>\epsilon \bigg)=0 ,
\label{eq:LLN}
\end{eqnarray}
for each $x\in \mathfrak{h}$,  $t>0$ and $\epsilon >0$.
\end{definition}

\begin{theorem}
Let $H$ be a random variable with values in the set of self-adjoint operators in a space $\mathfrak{h}$ that satisfies the conditions of Theorem 2. Let $( H_n )$ be a sequential identity of independent identically distributed random Hamiltonians, the distribution of each of which coincides with the distribution of the random Hamiltonian $\xi$. Then, for successive compositions, $\{ e^{i\frac{t}{n} H_n}\circ ...\circ e^{i\frac{t}{n} H_1},\, t\geq 0\}$, of independent identically distributed random semigroups, the law of large numbers holds in the strong operator topology (\ref{eq:LLN}).
\end{theorem}

Let us now investigate topologies on the space of measures on Banach spaces defined according to the following scheme. Given a mapping $\Psi$ that associates a measure $\mathbb{P}$ on a Banach space $E$ with a linear convolution operator $\Phi_\mathbb{P} \in \mathcal{L} (X)$ with a measure $\mathbb{P}$ of an arbitrary function $u$ from some topological vector space $X$ of measurable numerical functions on the space $E$:
\begin{eqnarray}
\Phi _{\mathbb{P}} \, u(x)=( \mathbb{E} S_{\xi })u(x)=\int\limits_Eu(x-y)\mathbb{P} (d y),\ u\in X.
\end{eqnarray}

The mapping $\Psi$ induces a topology on the space of measures on the space $E$ in some topology on the space of linear operators $\mathcal{L} (X)$ acting in the space $X$.

Let $E$ be a \emph{real} separable Hilbert space, and $\mathscr{B} (E)$ be the $\sigma$-algebra of Borel subsets of E. We denote by $\mathrm{BV} (E,\mathscr{B}(E))$ the Banach space of Borel measures of bounded variation on the measurable space $(E,\mathscr{B}(E))$. Let $X$ be some locally convex space of functions, $u:\ E\to \mathbb C$, and ${\cal L}(X)$ be a locally convex space of linear bounded operators in the space $X$.

\begin{theorem}
Weak convergence of a sequence of Borel measures $( \mathbb{P}_n )_n$ on the space $E$ is equivalent to the pointwise convergence of operators in a topological vector space $(C_b(E),\tau )$, where $\tau$ is the topology of pointwise convergence on $E$.
\end{theorem}

Let $E=\mathbb{R}^d$, $X=C_b(\mathbb{R}^d)$ endowed with the topology of pointwise convergence. A sequence $(\mathbb{P} _n)_n$ of Borel measures on a space $E$ with bounded variation is said to be weakly converging to the measure $\mathbb{P} \in \mathrm{BV} (E,\mathscr{B}(E))$ if for each function $f\in C_b(E)$ the equality $\lim\limits_{n\to \infty }\int\limits_Ef(x)\mathbb{P} _n(dx)=\int\limits_{E}f(x)\mathbb{P} (dx)$ holds true.
This condition is equivalent to the fact that, for each $f\in C_b(E)$ and $h\in E$, the equality $\lim\limits_{n\to \infty }\int\limits_Ef(x-h) \mathbb{P} _n(dh)=\int\limits_Ef(x-h)\mathbb{P} (dh)$ holds.
This means that a sequence of linear operators $\{ \Psi (\mathbb{P} _n)\}$ acting in the topological vector space $(C_b(E,\tau )$ converges pointwise to the operator ${\Psi }(\mathbb{P} )$.

\begin{definition}
A sequence of measures $\{ \mathbb{P} _n\} :\: {\mathbb N}\to \mathrm{BV} (E,\mathscr{B}(E))$ is called converging $\mathcal{L} (X)$-weakly to measure $\mathbb{P} \in \mathrm{BV}  (E,\mathscr{B}(E))$ if the sequence of operators $\Phi _{\mathbb{P} _n}$ acting in the space $X$ according to the rule $\Phi _{\mathbb{P} _n}u(x)=\int\limits_{E}u(x-y) \mathbb{P} _n(dy),\ u\in X,\, x\in E$, for $x \in E$, converges in the space ${\cal L}(X)$ to the operator $\Phi _{\mathbb{P} }$ correspondingly defined $\Phi _{\mathbb{P} }u(x)=\int\limits_{E}u(x-y)\mathbb{P} (dy),\ u\in X,\, x\in E$.
\end{definition}

\begin{definition}
A sequence of random variables $\{ \xi_n \}$ with values in the space $E$ is said to converge in distribution $\mathcal{L}(X)$-weakly to a random variable $\xi$ if the sequence of Borel measures $\mathbb{P}_n$ defined by the equalities $\mathbb{P} _n(A)=P(\xi _n^{-1}(A)),\ A\in \mathscr{B}(E),\, n\in {\mathbb N}$, converges $\mathcal{L}(X)$-weakly to the measure defined by the equality $\mathbb{P} (A)=P(\xi ^{-1}(A)),\ A\in \mathscr{B}(E)$. 
\end{definition}

If the topology $\tau_X$ of the space $X$ is given by a family of seminorms, $\{ \phi _{a},\, a\in A\}$, and the topology in the space of bounded linear operators $B(X)$ is strong operator, then the $(B(X),\tau _{sot})$-weak topology on the space of measures $\mathrm{BV}  (E,\mathscr{B}(E))$is given by the family of functionals, $\{ \Phi _{a,u},\, a\in A,\, u\in X\}$, defined by the equalities
\begin{eqnarray}
\Phi _{a,u}(\mathbb{P} )=\phi _a(\int\limits_{E}u(x-y) \mathbb{P} (dy)),\, \mathbb{P} \in \mathrm{BV} (E,\mathscr{B}(E)).
\end{eqnarray}

\begin{theorem}
\label{thm:main}
Let $\xi (t),\ t\geq 0,$ be a random process with values in space $E$, and let $(\xi _n )_n$ be a sequence of independent random processes, identically distributed with the common distribution of the process $\xi$. Let $X$ be a Banach space of functions such that for any function $u:\ E\to \mathbb C$ and for $t\geq 0$ we define the linear operator $u\to F(t)u= \mathbb{E}(S_{\xi (t)}u)$ is defined on the space $X$. Then if the operator function satisfies the conditions of Chernoff's theorem, then the sequence of random processes,$(\eta _n (t),\, t\geq 0 )_n$, where $\eta _n (t)=\xi _{n}(\frac{t}{n})+...+\xi _1(\frac{t}{n})$ converges in distribution $(B(X),\tau _{\mathrm{sot}})$-weakly to a Markov random process generating a semigroup, $\exp (F^\prime (0)t),\, t\geq 0$.
\end{theorem}

The identity $ \mathbb{E}(S_{\xi _n(t)}\circ ...\circ S_{\xi _1(t)}u)=(F(t))^nu$ is true for all $t\ge 0, n \in \mathbb{N}$, due to the independence of the processes $( \xi _n)_n $, see \cite{OSS19}. Since the function $F$ satisfies the conditions of Chernoff's theorem, for any we have $u\in X$ 
\begin{eqnarray}
\lim\limits_{n\to \infty }\sup\limits_{t\in [0,T]}\|  \mathbb{E}(S_{\xi _{n}(\frac{t}{n})+...+\xi _1(\frac{t}{n})} u)-\exp (F^\prime (0)t)u\| _X=0 \ \forall \ T>0.
\end{eqnarray}

This means $(B(X),\tau _{\mathrm{sot}})$-weak convergence in distribution of a sequence of random processes $( \eta _n (t),\, t\geq 0)_n $ to a Markov random process generating the semigroup $\exp (F'(0)t),\, t\geq 0.$ 

\begin{remark}
For a sequence of compositions of independent random shift operators by a vector of a Euclidean space, the assertion of Theorem \ref{thm:main} includes the assertion of the central limit theorem for sums of independent random vectors if we choose the space $C_b (E)$ as $X$ with the topology of pointwise convergence.
\end{remark}

\section*{Acknowledgments}
John Gough acknowledges funding under the French National Research Agency (ANR) grant Q-COAST ANR 19-CE48-0003.

\end{document}